\documentclass[fleqn,usenatbib]{mnras}

\usepackage{newtxtext,newtxmath}
\usepackage[T1]{fontenc}

\DeclareRobustCommand{\VAN}[3]{#2}
\let\VANthebibliography\thebibliography
\def\thebibliography{\DeclareRobustCommand{\VAN}[3]{##3}\VANthebibliography}

\usepackage{graphicx}	% Including figure files
\usepackage{amsmath}	% Advanced maths commands
\title{Search for periodic emission of five gamma-ray pulsars at the frequency of 111 MHz}

\author[S. A. Tyul'bashev et al.]{
Sergei A. Tyul'bashev,$^{1}$\thanks{E-mail: serg@prao.ru (SAT)}
Marina A. Kitaeva,$^{1}$
and Gayane E. Tyul'basheva$^{2}$
\\
$^{1}$ P.N. Lebedev Physical Institute of the Russian Academy of Sciences, Astro Space Center, Pushchino Radio Astronomy Observatory,\\
Radiotelescopnaya 1a, Moscow reg., Pushchino, 142290, Russia \\
$^{2}$ Institute of Mathematical Problems of Biology, brunch of Keldysh Institute of Applied Mathematics,\\  
Vitkevich 1, Moscow reg., Pushchino, 142290, Russia\\
}

\date{25.06.2021}

\pubyear{25.06.2021}

\begin{document}
\label{firstpage}
\pagerange{\pageref{firstpage}--\pageref{lastpage}}
\maketitle

\begin{abstract}
The search for pulsar (periodic) emission of five gamma-ray pulsars was carried out using the summed power spectra and the summed periodograms. No harmonics corresponding to the known pulsar periods were found. The upper estimations of the integral flux density of the pulsars J0357+3205 ($<$ 0.5 mJy), J0554+3107 ($<$ 0.5 mJy), J1958+2846 ($<$ 0.5 mJy), J2021+4026 ($<$ 0.4 mJy), and J2055+2539 ($<$ 0.55 mJy) is obtained.

\end{abstract}


%%%%%%%%%%%%%%%%%%%%%%%%%%%%%%%%%%%%%%%%%%%%%%%%%%

%%%%%%%%%%%%%%%%% BODY OF PAPER %%%%%%%%%%%%%%%%%%

\section{Introduction}
In the 74 years since the discovery of the first pulsar (\citeauthor{Hewish1968}, \citeyear{Hewish1968}), almost 3,000 pulsars have been discovered (https://www.atnf.csiro.au/research/pulsar/psrcat/). The vast majority of these sources are found in the radio range. In 1970, a paper was published talking about the probable detection of pulsating gamma emission in a young pulsar located in the Crab nebula (\citeauthor{Vasseur1970}, \citeyear{Vasseur1970}).

Until 2008, gamma emission was detected in pulsars that were initially found in other ranges. In 2009, the first works appeared with observations of samples of gamma-ray sources on the Fermi (\citeauthor{Abdo2009}, \citeyear{Abdo2009}) orbital telescope. It turned out that many of the previously discovered gamma sources are gamma pulsars. The vast majority of these pulsars have no emission in the radio range. It is customary to use the term "radio quite pulsars" for such pulsars. It is assumed that they have a thermal emission mechanism, unlike radio pulsars, whose emission is associated with the movement of electrons in strong electric and magnetic fields.

For several gamma-ray pulsars considered radio-quite, radiation in the radio range at wavelengths from centimeters to decameters (\citeauthor{Malofeev2005} (\citeyear{Malofeev2005}), \citeauthor{Camilo2009} (\citeyear{Camilo2009}), \citeauthor{Maan2014} (\citeyear{Maan2014}), \citeauthor{PeiWang2018} (\citeyear{PeiWang2018}), \citeauthor{Tyulbashev2019} (\citeyear{Tyulbashev2019})) was detected. In particular, in the work \citeauthor{Tyulbashev2019} (\citeyear{Tyulbashev2019}), the search for periodic radiation of the pulsar J0357+3205 was carried out in daily observations at an interval of 5 years, and the pulsar was detected in separate sessions only once. It is possible that the luminosity of gamma pulsars in the radio range is very low, and the sensitivity of modern radio telescopes is insufficient for their regular detection.

In this paper, an attempt is made to detect weak periodic radio emission of several gamma pulsars in observations on the radio telescope Large Phased Array (LPA) of the Lebedev Physical Institute of the Russian Academy of Sciences (LPI). Monitoring observations on the radio telescope have been going around the clock for six years. Approximately five days of observations have been accumulated in the direction of each LPA beam. Summation of power spectra and summation of periodograms are used to increase the sensitivity of observations, and a new visualization representation of the processed data is used to detect periodic radiation.

\section{Observations and processing program}

After the reconstruction of the LPA LPI, which ended in 2012, its effective area increased 2-3 times and is approximately equal to 45000-50000 sq.m. in the direction of zenith. The central frequency of observations is 110.3 MHz, and the full band of observations is 2.5 MHz. Based on the antenna field consisting of 16384 dipoles, several independent radio telescopes have been created. One of them is used for round-the-clock monitoring observations. This radio telescope has 128 stationary beams, which overlap in the plane of the meridian declination from $-9^o$ to $+55^o$. Until the end of 2020, observations were made in 96 spatial beams, overlapping declinations from $-9^o$ to $+42^o$. Since the beginning of 2021, 24 more spatial beams have been added in the test mode, and monitoring observations are now being carried out for declinations up to $+52^o$.

The number of beams serviced by one recorder is related to its capabilities for digitizing the data stream at the input. One recorder is used for 48 beams and the total input data stream is approximately 12 gigabits per second. The readout time generated by the registrar and the number of frequency channels are related to the physical capabilities of the RAM of the industrial computer used and the speed of writing information to the hard disk. Monitoring observations use a mode when the band is divided into 32 frequency channels with a width of 78 kHz, and recording is conducted with a readout time of 12.5 ms. Even for observations of second pulsars, this mode is not optimal. In the meter wavelength range, for most of the known pulsar tasks, it is preferable to have frequency channels with a width of 5-20 kHz and a readout time of 1-3 msec. However, in the non-optimal mode used, 35 terabytes of data are recorded in 96 beams per year. In the optimal mode, it would be necessary to record up to 6 petabytes per year, which exceeds the observatory's data storage capabilities.

For more information about the reconstruction of the meridian radio telescope of the LPA LPI and about the monitoring scientific programs running on it, see \citeauthor{Shishov2016} (\citeyear{Shishov2016}), \citeauthor{Tyulbashev2016} (\citeyear{Tyulbashev2016}). Additional details about the registrar can be found in the works \citeauthor{Tyulbashev2016} (\citeyear{Tyulbashev2016}), \citeauthor{Logvinenko2020} (\citeyear{Logvinenko2020}).

As is known, the sensitivity of observations on a radio telescope is determined by the temperature of the system, the effective area of the antenna, the receiving band, the readout time and the observation time. For LPA LPI, the observation time in one session is determined by the time of passage of the studied source through the meridian and is approximately 3.5 minutes at the equator at half power. According to \citeauthor{Tyulbashev2016} (\citeyear{Tyulbashev2016}), the average sensitivity at the zenith for observations of second pulsars outside the Galactic plane is 6-8 mJy, and in the Galactic plane 15-20 mJy. The best and worst sensitivity may differ from these values by about 1.5 times due to the fact that the coordinates of the sources do not coincide with the coordinates of the beams, and therefore not the full flux is observed in the beam, but only part of it. There are other corrections that take into account the features of the antenna array of the LPA LPI.

Sensitivity can be increased by increasing the observation time. For example, if the period and the derivative of the pulsar period are known with high accuracy, then it is possible to sum up the pulsar pulses observed on different days, months and years. Unfortunately, although the recorders on monitoring observations are started according to the atomic frequency standard, after starting, the time is counted by a quartz oscillator, which at a time interval of one hour gives a possible time error of $\pm 25$~ms. Attempts to do timing have so far been unsuccessful.

In the works \citeauthor{Tyulbashev2017} (\citeyear{Tyulbashev2017}), \citeauthor{Tyulbashev2020} (\citeyear{Tyulbashev2020}) on the search for second pulsars at the LPA LPI, it was proposed to increase the sensitivity of observations due to incoherent addition of power spectra. Information about the pulse phase is lost in the power spectrum, but the location of harmonics in the spectrum for a given pulsar is the same, regardless of the day of observations. If we approach strictly, then, of course, the pulsar period changes with time, but since the accuracy of the period determined by the power spectrum in a 3.5 minute recording is no better than one in the third decimal place, then we will not feel changes in the pulsar period at intervals of hundreds of years. The harmonics of pulsars will always fall on the same numbers of points in the power spectrum. By obtaining power spectra for different days of observations and adding them together, it is possible to increase the signal-to-noise ratio (S/N) of harmonics observed in the spectrum. The sensitivity should increase as the square root of the number of stacked spectra if the original noise was white and the antenna gain was unchanged over the entire observation interval. However, the sensitivity may vary slightly from day to day due to the different physical condition of the antenna and weather conditions. Note also that not all interference can be removed during processing. For these reasons, the actual sensitivity increases to a lesser extent than expected. To calculate the estimate of the increase in real sensitivity in the summed power spectra, we obtain independent estimates of the magnitude of the initial noise and the increase in sensitivity for each direction in the sky (details in \citeauthor{Tyulbashev2020} (\citeyear{Tyulbashev2020})).

As is known, the search for pulsars can also be carried out using periodograms. According to the works of \citeauthor{Cameron2017} (\citeyear{Cameron2017}), \citeauthor{Parent2018} (\citeyear{Parent2018}), \citeauthor{Morello2020} (\citeyear{Morello2020}), the sensitivity when searching using periodograms may be higher than when searching using power spectra. In the program for processing monitoring data of the LPA LPI, both methods of searching for periodic radiation are implemented.

There are common standards for finding new pulsars. In search programs, power spectra or periodograms are first constructed, and when sorting through possible measures of dispersion (DM), harmonics are searched for whose S/N level is greater than the set value. According to various criteria, false sources are eliminated, and the remaining candidates are viewed visually. For visual viewing, images are created that show the resulting average profile, the dynamic spectrum, and the dependence of the peak flux density in units of S/N on DM. For sources that have passed visual verification, additional observations are carried out, and, if possible, their period and the derivative of the period are specified.

Such an observation processing scheme will work adequately if sources are observed whose flux density is such that they are visible in one observation session. As shown in \citeauthor{Tyulbashev2020} (\citeyear{Tyulbashev2020}), when searching for very weak pulsars, situations arise when harmonics with S/N$>$7 are observed in the summed power spectra, including for pulsars detected in observations on other telescopes, but there is not a single session when we can get the average pulsar profile.

A new program for processing and visualizing the processed data has been created to search for pulsars that do not detect radiation in individual sessions. The right ascension and declination of pulsars according to the catalog for the year 2000 are used as input parameters. The program recalculates the coordinates for a given day and evaluates the quality of the noise track at the location of the pulsar. If the quality of the noise track is low, then this day is not involved in further work. If the quality of observations is good, then a piece of recording with a length of 16384 points (approximately 204.8 seconds) is cut out and power spectra are plotted when sorting through the dispersion measures from 0 to 1000 pc/cm$^3$. An additional search is also performed, taking into account that the pulse width of the pulsar may be greater than one point of the original (raw) data. To do this, the addition of 2, 4 and so on points in the raw data is performed, and the power spectra are built again. There are 6 such iterations in total and they allow us to obtain the maximum ratio of S/N in the power spectrum with an estimated pulse width from 12.5 ms to 800 ms. For each DM to be tested and for each iteration taking into account the pulse width, the corresponding power spectra are added up for all days of observations. In each summed power spectrum, the S/N of each point is determined and tables are created in which the values of the amplitudes of harmonics with S/N $>$4 are stored. These tables are used for visualization when searching for new pulsars.

\begin{figure*}
\begin{center}
	\includegraphics[width=1.0\textwidth]{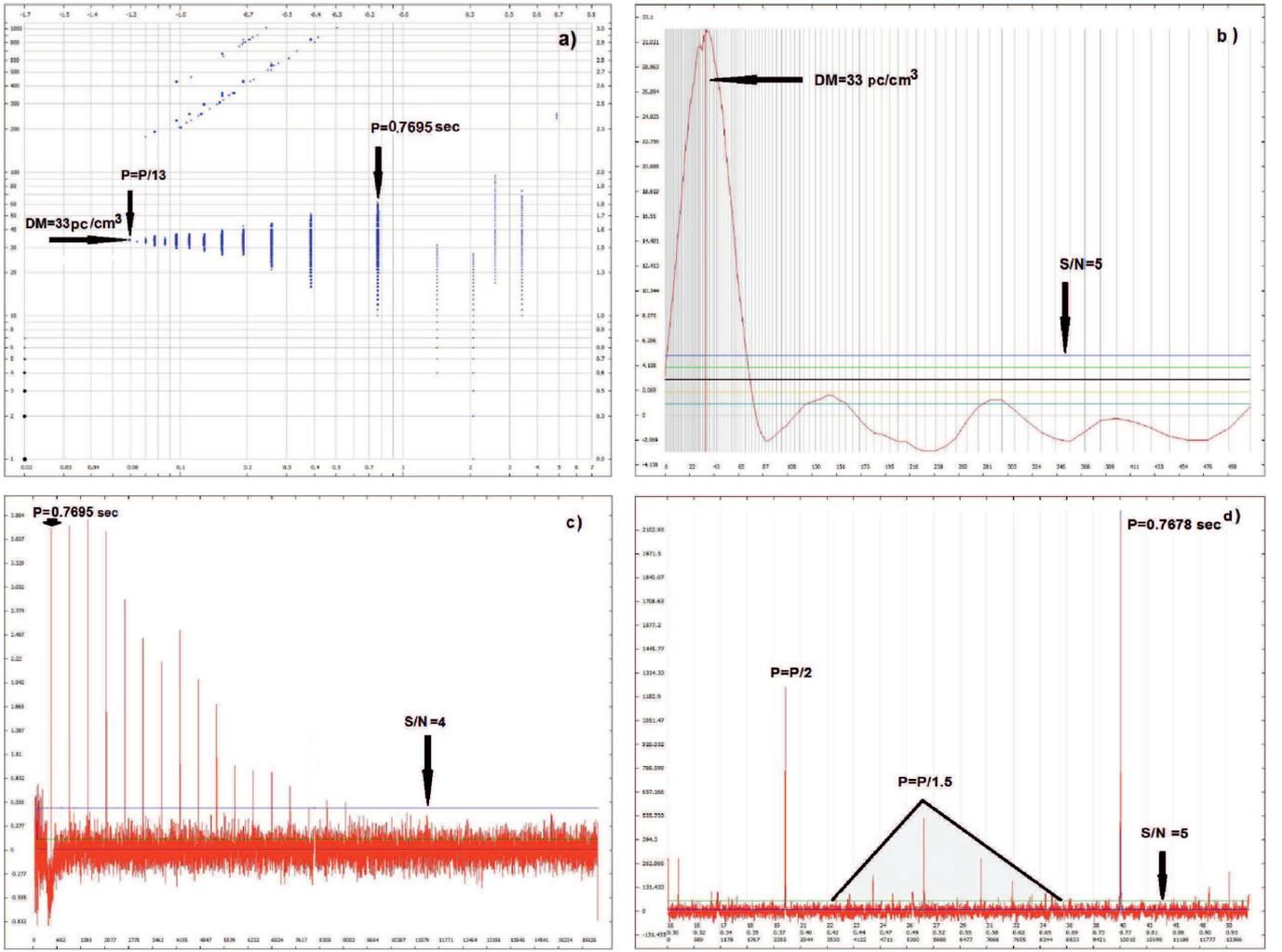}
    \caption{Figure 1a-d. Figure 1a shows a fragment of the main working window. On the OX axis, the time (period) in seconds, on the OY axis, the DM in pc/cm$^3$. The upper horizontal and right vertical scales reflect the values of the OX and OY axes in logarithmic format. Fig.1b shows a fragment of the working window reflecting the dependence of the height of the first harmonic in the summed power spectrum in units of S/N (OY axis) from DM (OX axis). Figure 1c shows the power spectrum summed up over a 3-year interval, assuming that DM= 33 pc/cm$^3$. The OX axis shows the numbers of points in the power spectrum. The signal power in conventional units is marked along the OY axis. Figure 1d shows a periodogram summarized over an interval of 5 years. Along the OY axis, the signal power in conventional units. There are three different digitizations along the OX axis. The upper one shows the "foldbins" parameter (\citeauthor{Morello2020} (\citeyear{Morello2020})), the middle digitization shows periods, the lower one shows the numbers of points in the periodogram. The presented periodogram is calculated for DM=34 pc/cm$^3$. The horizontal lines in Fig.1b-d mark different levels of the S/N, which allows you to navigate when viewing data.}
    \label{fig:fig1_gamma_eng}
\end{center}
\end{figure*}

In the central window of the visualization program (Fig.~\ref{fig:fig1_gamma_eng}), you can see a map where the location of the circle on the abscissa axis reflects the pulsar period (P), and on the ordinate axis the observed DM. The size of the circle reflects the S/N harmonics in the summed power spectrum. It is obvious that the maximum S/N in the harmonic will be observed in the power spectrum, which was calculated after adding the frequency channels with a pulsar DM and averaging the initial data, which correspond to the width of the average pulsar profile. However, a strong pulsar will be visible both on measures of dispersion close to the true one and on averages that do not coincide with the pulse width of the pulsar. Therefore, on the P/DM map, the pulsar should be observed in the form of vertical stripes narrowing to the edges and limited in height. The maximum size of the circle on this strip should be on the true DM. Smaller strips may appear on multiple harmonics corresponding to half of the period, third of the period, and so on.

The map generated by the data visualization program is interactive. By clicking the mouse on the circle of interest, you can view the power spectrum corresponding to the selected circle, build a dependence of S/N on DM for the selected harmonic. Thus, the program allows you to see the pulsar on the P/DM map, evaluate its DM, the harmonic shift and the expected pulse width. Some additional details concerning the processing and visualization program can be found in the work on the search for weak pulsars in the monitoring survey at the LPA LPI  (\citeauthor{Tyulbashev2021}, \citeyear{Tyulbashev2021}). An analog of the search program and visualization of the processed data are also made for searching using periodograms. The subroutine for constructing periodograms is taken from  \citeauthor{Morello2020} (\citeyear{Morello2020}) ( https://github.com/v-morello/riptide )

Fig.~\ref{fig:fig1_gamma_eng} shows an example of visualization when processing observations of the well-known pulsar J1638+4005 at an interval of 3 years for the summed power spectrum and at an interval of 5.5 years for the summed periodograms. Pulsar J1638+4005, having P=0.76772 s and DM=33.4 pc/cm$^3$ (\citeauthor{Tan2020}, \citeyear{Tan2020}), was detected in observations at the LPA LPI (\citeauthor{Tyulbashev2017}, \citeyear{Tyulbashev2017}) in the summed power spectra, and was noted in the original work as a weak pulsar. The average pulsar profile given in the article, obtained in one observation session lasting 3.5 minutes, shows S/N=6. In \citeauthor{Tan2020} (\citeyear{Tan2020}), estimates of the integral flux density of this pulsar are given from observations at LOFAR and on the 76-meter Jodrel-Bank telescope: 128 MHz (S=3.1 mJy), 167 MHz (S=1.7 mJy), 334 MHz (S=0.34 mJy), 1532 MHz (S$<$0.06 mJy). Based on the estimate of the spectral index given in the paper ($\alpha =2.3, S\sim \nu^{-\alpha}$), it is possible to estimate the expected integral flux density of this pulsar at the central frequency of the antenna of the LPA LPI ($S_{110.3MHz} =5.7$~mJy). The flux density of 5.7 mJy is very close to the best sensitivity of the LPA in one observation session (\citeauthor{Tyulbashev2016}, \citeyear{Tyulbashev2016}).

Fig.~\ref{fig:fig1_gamma_eng} shows that the pulsar J1638+4005 has 13 harmonics with S/N$\ge 6$. All signals with S/N $\ge 10$ are represented by circles of the same size, and therefore the width of the vertical segments varies slightly. The rightmost strip corresponds to a period of 0.7695 s (X axis), and the center of the strip corresponds to DM= 33-34 pc/cm$^3$ (Y axis). Depending on the (S/N)/DM for the first observed harmonic, the maximum falls on DM =33-34 pc/cm$^3$. The pulsar period, determined by point 532 in the power spectrum, is 0.7695 s. It does not match the catalog value of the period 0.7677 s. The difference in periods is due to the low frequency resolution in the power spectrum. We estimate the accuracy of the period, determined by the power spectrum, as 0.001s. When working with the visualization program, you can select any point on the power spectrum and see what period this point corresponds to, as well as from this point. If we add up the heights of all harmonics visible in the power spectrum, then the S/N of such a summed harmonic will be approximately 70. Thus, the pulsar detected in individual sessions at the sensitivity limit of the LPA LPI radio telescope is visible as a strong object in the summed power spectrum. It is also obvious that if this pulsar had a slightly lower flux density, it would not have been detected in separate observation sessions at the LPA, but would have been easily detected by a search program, even if it was 10 times weaker, and its expected integral flux density would have been about 0.5 mJy.

According to the works noted above, \citeauthor{Cameron2017} (\citeyear{Cameron2017}), \citeauthor{Parent2018} (\citeyear{Parent2018}), \citeauthor{Morello2020} (\citeyear{Morello2020}), periodograms can detect weaker pulsars than pulsars detected by power spectra. This is especially noticeable for pulsars, whose pulses are very narrow in relation to the pulsar period. There is also a gain when using periodograms when searching for pulsars with long periods. It is known that low-frequency noise is observed in the power spectra, which is difficult to subtract. An example of such noise at the beginning of recording is seen in Fig.1b. The periodogram in Fig.1d is obtained by summing up 5 years of observations (approximately 1800 observation sessions of 3.5 minutes each). The maxima marked with arrows on the periodogram correspond to the pulsar period P= 0.7678s and multiple periods. The light gray color in the figure shows the characteristic triangular structures that appear for strong pulsars when using periodograms. The peak marked in the figure has S/N=183. Based on the fact that the pulsar J1638+4005 was detected earlier at S/N=6 (\citeauthor{Tyulbashev2017} (\citeyear{Tyulbashev2017}), and its expected integral flux density is 5.7 mJy (see above in this paragraph), we can recalculate the minimum detectable signal in the summed periodogram and give an experimental estimate of the integral flux density of extremely weak pulsars $S=5.7/(183/6)=0.19$ mJy. This estimate is close to the theoretical estimate of the flux density of extremely weak to detect pulsars observed at the zenith exactly in the center of the beam, $S=5/(1800)^{1/2}=0.12$ mJy. Taking into account the typical loss of sensitivity by 1.5-2 times from the theoretical values due to weather conditions, lack of observations due to routine maintenance on the antenna, corrections for the features of the antenna array having a fixed direction of rays in the sky and other reasons, the estimates obtained are in good agreement. Thus, the program for searching for periodic signals using summed power spectra and summed periodograms can detect pulsars with a high degree of reliability, in which it is impossible to build an average profile based on observations in one session at the LPA LPI.

\section{Results and discussion}

The search for regular radiation was carried out for five gamma pulsars that got into the monitoring area. Pulsars J0357+3205, J0554+3107, J1958+2846, J2021+4026, J2055+2539 were detected in observations on the Fermi satellite (\citeauthor{Abdo2009} (\citeyear{Abdo2009}), \citeauthor{SazParkinson2010} (\citeyear{SazParkinson2010}), \citeauthor{Pletsch2013} (\citeyear{Pletsch2013})). Radio emission was detected only at the pulsar J0357+3205 in the observations of FAST/Arecibo (\citeauthor{PeiWang2018}, \citeyear{PeiWang2018}) and at the LPA LPI (\citeauthor{Tyulbashev2019}, \citeyear{Tyulbashev2019}). FAST/Arecibo observations are published in a presentation presented at the conference, and the details of detecting a periodic signal are practically absent. In the search for radio emission at the LPA LPI, a pulsar was detected once during 1700 observation sessions at S/N$>$7. The flux density in FAST/Arecibo observations at a frequency of 1250 MHz is 0.04 mJy, and at a frequency of 110.3 MHz – 14 mJy. 

The search for periodic radiation in the monitoring data of the LPA LPI was carried out in summed power spectra and in summed periodograms in observations over an interval of 5.5 years. No point thickenings, that is, points located along the line at known pulsar periods and related to close DM, were found for any pulsar. In contrast to the one shown in Fig.1a, the search maps show signals with S/N$>$4. Therefore, even if harmonics with S/N=4-5 were observed in the gamma pulsars under study, small vertical segments would have to be observed on some measures of dispersion when a periodic signal was detected. The absence of harmonics corresponding to pulsar periods makes it possible to obtain an upper estimate of the flux density under the assumption that the pulsar has a constant, albeit very weak, radiation in the radio range. Processing shows that for none of the five pulsars studied there are no signs of periodic signals with S/N$>$4.

Since we know the exact coordinates of the pulsars, we can calculate their location in relation to the fixed beams of the LPA and make corrections that allow us to obtain upper estimates of the flux density taking into account the features of the antenna array.

\begin{table}
	\centering
	\caption{Upper estimates of the integral flux density of the gamma pulsars}
	\label{tab:tab1}
	\begin{tabular}{ccp{1cm}cc} % four columns, alignment for each
		\hline
name & P (s) & increase of SNR & correction & $S_{int}$~(mJy)\\
		\hline
J0357+3205 & 0.44410  & 32.1     & 0.30     & $<0.5$ \\ 
J0554+3107 & 0.46496  & 27.5     & 0.70     & $<0.5$ \\
J1958+2846 & 0.29040  & 25.6     & 0.80     & $<0.5$ \\
J2021+4026 & 0.26532  & 34.3     & 0.70     & $<0.4$ \\
J2055+2539 & 0.31956  & 30.4     & 0.60     & $<0.55$ \\
		\hline
	\end{tabular}
	\label{tab:tab1}
\end{table}

Consider the processing of observations using the example of the pulsar J0357+3205. In the gamma range, the main component at half power takes a quarter of the period  (\citeauthor{Abdo2009} (\citeyear{Abdo2009}), i.e. about 100 ms. In the radio band at a frequency of 1250 MHz, the profile is two-component (\citeauthor{PeiWang2018}, \citeyear{PeiWang2018}). Based on the profile in the figure, one of the components is wide and its half-width (We) is approximately 90 ms, the second component is narrow, has a comparable height, and its half-width is approximately 35 ms. The distance between the components is 165 ms. Expected distance to the pulsar, 270-900 ps  (\citeauthor{Kirichenko2014}, \citeyear{Kirichenko2014}). According to observations, FAST DM=47 pc/cm$^3$. In the observations on the LPA, only one narrow component with DM=46-48 pc/cm$^3$ is visible, having S/N=10 (\citeauthor{Tyulbashev2019}, \citeyear{Tyulbashev2019}). In the average profile in the gamma range, the narrow component visible in the radio range is also guessed, but its height is significantly less than that of the wide component. 

The expected increase in sensitivity in the summed power spectra and in the summed periodograms is proportional to the square root of the number of observation sessions. For 5.5 years, the pulsar should be observed almost 2000 times. However, as mentioned in the previous paragraph, some observations disappear due to strong interference, there are days when there were no observations due to technical work on the antenna. During the remaining days, the background noise signal may change due to weather conditions. In order to obtain an estimate of the real growth of S/N, depending on the number of observation sessions, the noise dispersion was estimated each day at the time interval corresponding to the passage of the pulsar through the meridian. All noise variances were arranged in ascending order and normalized to the minimum variance. Thus, the minimum variance turned out to be equal to one, and the total variance is equal to the square root of the sum of the squares of the individual variances  (\citeauthor{Tyulbashev2020}, \citeyear{Tyulbashev2020}). The change in the S/N ratio depends on how many individual power spectra were formed and what normalized noise dispersions they had. 1334 sessions were used to add up individual power spectra. The theoretical value of the growth of the S/N should be $1334 ^{1/2} = 36.5$ times, the real growth of the S/N from the experiment is 32.1 times.

No outstanding details are observed at the location of the harmonic J0357+3205. The absence of a signal in the summed power spectrum allows us to give us an upper estimate of the integral density of the pulsar flux. Since the LPA LPI is an antenna array with fixed declination rays formed using the Butler matrix, a number of corrections must be made to estimate the flux density, taking into account the antenna features. These corrections are due to the fact that the pulsar coordinate does not coincide with the beam location, the pulsar is not observed at the zenith, and, therefore, the effective area of the antenna is less than 45,000 sq.m., each Butler matrix forms 8 rays, and they have a common envelope. For the pulsar J0357+3205, these three corrections give a coefficient of 0.3. That is, for this pulsar, only a third of the energy coming from the sky is observed on the LPA antenna. Assuming that the minimum noise dispersion determines the "ideal" observations and based on the sensitivity estimate of 5 mJy when observing pulsars located outside the galactic plane in the direction to the zenith (\citeauthor{Tyulbashev2016}, \citeyear{Tyulbashev2016}), we can give an upper estimate of the pulsar flux density: $S_{110.3MHz} < 5/(32.1\times 0.3) < 0.5$ mJy. The obtained estimate says that the integral flux density over a long-term observation interval is less than 0.5 mJy, but cannot guarantee that there were no periods of short flare activity during this observation interval.

The same analysis was done for the remaining four pulsars as for J0357+3205. Harmonics at known periods were also not detected for all pulsars. The distances to pulsars were estimated indirectly by different authors. Usually, the authors proceeded from the assumption that pulsars are young and should be located near the remnants of supernovae that gave rise to pulsars. Estimated distances to the remaining gamma pulsars: 3.5 kps (J0554+3107; \citeauthor{Pletsch2013} (\citeyear{Pletsch2013}), 9.2 kps (J1958+2846; \citeauthor{Aleksic2010} (\citeyear{Aleksic2010}), 2 kps (J2021+4026; \citeauthor{Hui2015} (\citeyear{Hui2015}), 0.6 kps (J2055+2539; \citeauthor{Mignani2018} (\citeyear{Mignani2018}). Unlike pulsar J0357+3205, pulsars J0554+3107, J1958+2846, J2021+4026 and J2055+2539 lie in the plane of the Galaxy at galactic latitudes not exceeding $10^o$. Therefore, when obtaining the upper estimate of the flux density, it was assumed that in the direction to the zenith, the minimum detectable integral flux density in a single observation session for these four pulsars was not 5, but 10 mJy (\citeauthor{Tyulbashev2016}, \citeyear{Tyulbashev2016}).

Having done the same work for the remaining pulsars as for J0358+3205, we obtained upper estimates of the integral flux density for all sample sources. The results are shown in Table 1. In the first column of this table – the name of the pulsar in the J2000 annotation, in the second – its period, in the third – the expected gain of the S/N in the accumulated power spectra and periodograms, in the fourth – multiplied corrections taking into account the loss of the signal. In the fifth column, an upper estimate of the integral flux density of the pulsar at a frequency of 110.3 MHz is given, assuming that the pulse broadening inside the frequency channel due to the measure of dispersion is insignificant.

The absence of harmonics in the averaged power spectra and on periodograms may be due to several reasons. Firstly, the level of the integral flux density may be lower than the upper estimate obtained. Secondly, when obtaining the upper estimate of the flux density, factors related to the peculiarities of observations on a antenna array having a fixed location of beams in the sky were taken into account, and then the sensitivity was estimated according to the standard radiometric gain formula. If there are any additional unaccounted factors that reduce sensitivity when searching for new pulsars, then the upper estimate of the flux density may increase. Thirdly, the obtained estimates of the flux density were made under the assumption that the DM does not introduce additional broadening of the pulse inside the frequency channel. Based on the width of the frequency channel, it is easy to calculate that the intra-channel broadening of the pulse by DM = 100 pc/cm $ ^3$ doubles the deterioration shown in the Table~\ref{tab:tab1} ratings. Fourth, it was assumed that the pulse duration is $<$~0.1 of the period. If the pulse width in the radio range is equal to half of the period, then the upper estimates of the flux density in Table 1 will change twice and amount to approximately 1 mJy for all five pulsars. Fifthly, if the radiation of gamma pulsars is flashy in the radio range, or they have a strong variability associated with other reasons, then there may be a situation when the integral flux density determined at the full observation interval is less than 0.5 mJy, and at certain time intervals the pulsar is still visible.

% Don't change these lines
\bsp	% typesetting comment
\label{lastpage}
\end{document}